\documentclass[openacc]{rstransa}
\usepackage{epsf}
\usepackage{tikz}

\titlehead{Research}

\begin{document}

\title{Quantum Haecceity}

\author{Ruth E. Kastner$^{1}$}

\address{$^{1}$University of Maryland, College Park\\}

\subject{xxxxx, xxxxx, xxxx}


\corres{R. E. Kastner\\
\email{rkastner@umd.edu}}

\begin{abstract}
There is an extensive philosophical literature on the interrelated issues of identity, individuality, and distinguishability in quantum systems. A key consideration is whether quantum systems are subject to a strong form of individuality termed ``haecceity'' (from the Latin for ``this-ness'').  I argue that the traditional, strong form of haecceity does not apply at the quantum level, but that in order to properly account for the need for symmetrization in quantum systems, a weaker kind of haecceity must be involved, which I call {\it quantum haecceity}. In the process, I also question some generally accepted tenets of the current debate, such as the idea that symmetrization of states for identical quanta must be postulated and reflects permutation invariance. Instead, I note that a perturbative Hamiltonian is required for exchange effects, which suggests that the need for symmetrization arises from specific physical conditions. 
\end{abstract}

\maketitle

\section{Introduction} 

\subsection{Background}  

There is a vast, centuries-old philosophical literature on the interrelated issues of identity, individuality, and distinguishability. An exemplar of this discussion is Leibniz'  Principle of the Identity of Indiscernibles (PII), which asserts that if there is no way to distinguish between two things on the basis of their properties, then they are the same thing. Quantum theory complicates this already complex and subtle set of metaphysical issues. In particular, there are quanta that have identical essential properties, but which cannot be identified as the same individual, which conflicts with the PII.  Symmetrized states (superpositions of mutually permuted direct product states) involve systems that are indistinguishable in their essential properties, but that nevertheless cannot be identified as the same entity, since their cardinality is greater than unity.  (The present author is in broad agreement with the view that due to this situation, quantum theory cannot be reconciled with the PII.) \smallskip


 In the context of quantum theory, an extensive literature spans various controversies and lack of consensus. Areas of contention are not limited to considerations of how best to resolve the challenges, but encompass matters of basic definition as well. For an overview of the debate, see, e.g. Ladyman and Bigaj (2010) and references therein. Other useful references are  French and Redhead (1988) and the more recent French (2019). In the current work, I will not address disagreements concerning basic definitions of the concepts, which concern subtleties of meaning that I believe are not relevant for my present purpose. I will focus on two specific issues  concerning symmetrized  states: \smallskip
 
\noindent (1) What is it that is physically (or metaphysically) exchanged when indices are exchanged?\newline
\noindent (2) How universal are symmetrized states, especially in regard to exchange forces?\smallskip

Before proceeding to address these questions, a bit more background concerning terminology is in order.  The term ``factorism'' has arisen in the context of this debate. It refers to the idea that quantum systems can be represented by direct products of individual Hilbert spaces, e.g. $|Up\rangle \otimes  |Down\rangle $ , or more compactly,  $|Up\rangle  |Down\rangle $  , in a physically meaningful way.  ``Factorism'' is eschewed and critiqued by various researchers active in this area---in particular, A. Caulton (2018), who introduced the term, T. Bigaj, (2022), and Dieks and Lubberdink (2022), and the term conveys a pejorative connotation. I believe that this repudiation of the direct product states is misguided. The present paper will introduce some reasons for questioning the repudiation of the direct product spaces, but it will not attempt to address that issue in comprehensive detail, since it merits a much more expansive treatment. Nevertheless, to adopt a more neutral tone towards what is termed ``factorism,''  I will use the term ``direct product states'' to denote such non-symmetrized states. 

 In particular, I will argue that the individual direct product state components of the symmetrized states are not only physically meaningful, but (addressing question (1) above) that their permutations represent a change in the physical situation that is properly understood in terms of an exchange of haecceities. However, the customary strong notion of haecceity as a transcendent, persistent form of individuality does not apply to quantum systems, since the identities of indistinguishable quantum systems do become blended in a way cannot be denied under pain of empirical falsification. Thus, I argue that what is needed is a specific weaker form of haecceity, which I call {\it quantum haecceity}. Along the way, (addressing question (2) above) I will also question what I will term the "Received View" that symmetrization must be considered an independent and universal postulate (as opposed to arising directly from physical theory under specific conditions). However, the present paper makes no pretense of providing an exhaustive treatment of the Symmetrization Postulate (SP). \smallskip

\subsection{Haecceity } 

We first briefly review the concept of haecceity. The term is based on the Latin ``haec'' (this), and roughly translates to ``thisness.'' It is used to denote a concept of property-independent individuality. Thus, it is a form of individuality that transcends all qualitative aspects of an entity, and confers on it a unique identity that distinguishes it from all other entities, regardless of properties. 

In this regard, we need to recall the distinction between essential properties and contingent properties. In doing so, let us use the rather old-fashioned example of a coin, simply because it is still an effective one. An essential property is what makes an object that type of object. So for example, essential properties of a penny are that it is made of a copper alloy and has Lincoln's bust embossed upon the ``heads'' side with a contrasting design on the ``tails'' side. In contrast, a contingent property is not an essential feature of a coin but one that it may possess under certain circumstances. An example is whether it landed ``heads up'' or ``tails up''  when tossed. Contingent properties correspond to states in physics.\footnote{The Spanish language nicely captures this distinction through its two different forms of the verb ``to be'': {\it ser} denotes an essential property, while {\it estar} denotes a contingent property.}  In the quantum context, typical observables corresponding to properties (i.e., eigenvalues and their associated eigenstates) are energy, momentum, angular momentum, and position (however the latter applies only approximately, at the non-relativistic level). Thus, to be clear, I take quantum states in a basic realist way: i.e. that quantum states describe physical systems in their capacity to instantiate specific contingent properties. However, unlike in classical physics, such properties may be only provisional and indeterminate (if a system is in a superposition of states with respect to a given observable).\footnote{So, for example, I would dissent from a typical characterization of quantum states like the following found in an online discussion : "the wave function isn't a `real object' that can be measured or seen by one experiment. It is a collection of numbers storing all the probabilities [for outcomes] of a measurement.'' (L. Motl, 2012). This is a statement of instrumentalism about quantum theory, together with an assumption that an entity's being real requires that the entity is empirically detectable. This is a widespread ``default'' position among researchers today. More formally, the Stanford Encyclopedia provides the following description of the instrumentalist stance: ``Traditionally, instrumentalism holds that claims about unobservable things have no literal meaning at all...Some antirealists contend that claims involving unobservables should not be interpreted literally, but as elliptical for corresponding claims about observables.'' (Chakravartty, A., 2017). Ernan McMullin counters this view with his comment: ``Imaginability must not be made the test for ontology. The realist claim is that the scientist is discovering the structures of the world; it is not required in addition that these structures be imaginable in the categories of the macroworld.'' McMullin, (1984). If the micro-structures are difficult even to imagine in the categories of the macroworld, clearly they need not be empirically observable. } \smallskip

Returning now to the classical situation: If we had two qualitatively identical coins (such as two ideal pennies), attributing haecceity to each of the coins through the use of labels such as ``Coin 1'' and ``Coin 2'' would mean that regardless of the degree of similarity of their properties, they are not the same coin. Thus, in general, haecceitism is in tension with Leibniz'  Principle of the Identity of Indiscernibles (cf. Loemker, 1969). Classically, we can get away with labeling of systems in this way without necessarily attributing haecceity to them, since classical systems are never absolutely indiscernible in the way that quantum systems are. That is, ``classical systems''\footnote{I use scare quotes because these are always idealizations; the world is quantum-mechanical at the micro-level. One may indeed find classical {\it phenomena}, but that does not mean that underlying such phenomena are things legitimately characterized as ``classical systems.''}  are viewed as intrinsically distinguishable in a way that quantum systems are not.  Thus, the labeling in classical physics need not be viewed as an attribution of haecceity, but rather as a reflection of the assumed intrinsic distinguishability of classical systems. Thus, haecceity is optional in the context of classical physics: it is not required, but can be applied consistently, and labels interpreted as indexing haecceities. \smallskip

The foregoing means that we can take the meaningful attribution of a label to a system as a necessary but not sufficient condition for haecceity of a system. Thus, attribution of a unique (non-interchangeable) label on a system means either (a) that system possesses haecceity or (b) that system is distinguishable, by virtue of its properties, from other systems. (In the latter case, the label is a surrogate for the distinguishing property.) For specificity and in preparation for what follows, I will call the traditional notion of haecceity pertaining to (a)  {\it classical haecceity}. I will propose that use of labels as interchangeable among a number $N$ of quanta with identical essential properties reflects a weakened form of haecceity that I will call {\it quantum haecceity}; what I mean by ``weakened'' is discussed in Section 4. On the other hand, I will argue in Section 5a that the second, distinguishability interpretation of labels (b) carries over into quantum theory in a physically significant way, in the context of measurement.
 
\subsection{ Classical vs quantum state spaces} 
 
Let us now return to our classical coin example, and take the labels as denoting haecceities. If we have more than one system in play, then we have a set of joint states constituting the collective state space. In this schema, for coins that are labeled ``1'' and ``2',' we have four possible overall configurations: \smallskip

\noindent (a) Both heads\newline
(b) Both tails\newline
(c) 1 heads, 2 tails\newline
(d) 1 tails, 2 heads\newline
\smallskip

 The key aspect that signals classicality here is that there is a fact of the matter about which coin is in which state for the heterogeneous states. \smallskip

In quantum theory, as is well known, the situation differs in a crucial way. For quanta that have the same essential attributes, for empirical consistency (under certain conditions, e.g. in statistical descriptions) we cannot make use of the heterogeneous states (c) and (d), in which the quanta occupy specific differing states as a ``fact of the matter.''  Instead, we must describe such multi-quantum systems with one collective heterogeneous state, with no fact of the matter about which quantum is in which state. Thus, (c) and (d) become amalgamated into what we may call (c'), and we have instead (where now the possible states are, for example, electron spin `up' or `down' with respect to some axis): 

\bigskip

\noindent (a) Both up \newline
(b) Both down\newline
(c') one up, the other down

\bigskip

Thus, we cannot say that either electron is definitely in either state. This is often thought of as signifying that there is no physical distinction between the cases \smallskip

 ``electron 1 up and electron 2 down''
 and  
`electron 1 down and electron 2 up.'' \smallskip

However, I will argue that it is not legitimate to conclude from ``there is no fact of the matter about which electron is in which state'' that the permuted states refer to a single physical situation (i.e., that there is a representational redundancy). For one thing, such an inference is arguably an invocation of the PII, which is already called into question by quantum theory. Interpretation of these kinds of permuted direct product states will be considered in more detail in what follows.

\subsection{Symmetrized states in quantum theory} 

As discussed above, for quanta with identical essential properties, one cannot (in general) attribute a particular state to either system (as in (c) and (d) above), and one must instead describe the system by the ``symmetrized'' state that results from permuting the labels and summing both in equal amplitudes:

$$ S \pm =  \frac{1}{\sqrt 2} ( | \uparrow_1\rangle | \downarrow_2 \rangle  \pm | \uparrow_2\rangle | \downarrow_1  \rangle)  \eqno(1) $$

\noindent  The state  (1) is the quantum theoretical description of the situation represented by (c') above. It represents two different direct product terms in a quantum superposition (where there are two possible signs for the relative phase between the terms for bosons and fermions). Thus, there is no fact of the matter about which system is in which state. We must represent the overall state as involving equal amounts of each heterogenous state, which is accomplished by exchanging the indices and constructing a superposition asserting that the original state and the permuted state must be equally present. This raises the following question: what does it mean physically (or at least metaphysically) to ``exchange the indices''?

T. Bigaj (2015, 2022) has suggested some possible ways of interpreting the physical content of the exchange of the indices, where two distinct interpretations emerge as most likely to be physically applicable. These are (i) exchange of essences (EE) and (ii) exchange of haecceities (EH). 
Option (i), exchange of essences (EE), involves the idea that when we permute labels or indices among systems, we are exchanging their essential qualities or properties. Since quantum systems of a given type, such as electrons, have identical essential qualities, exchanging the essences of the two systems changes nothing about the physical situation, and the result is taken as a redundancy of description; i.e., two ways of describing the same thing. Option (ii), exchange of haecceities (HE), involves the idea that exchanging the indices or labels exchanges the haecceities of the systems. Recalling that haecceity is a form of individuality that transcends properties, HE implies that the two states in the superposition (1) describe distinct physical situations. However, before dealing specifically with the choice of EE or EH as the appropriate one, we consider first a few concerns about the current state of debate.

\section{Symmetrization and exchange: some concerns} 

\subsection{Symmetrization {\it because of} invariance or {\it for} invariance?} 
	 	
In the literature, symmetrization is traditionally characterized as a ``postulate'' arising from ``exchange degeneracy.'' I will call the view that quantum theory requires a "Symmetrization Postulate" (SP) the "Received View." 
As noted in the Introduction, I recognize that the subject of the SP demands a deep and broad treatment and I make no pretense of doing  it justice in the present, narrowly focused work. However, it does seem appropriate to at least point to the fact that the Received View can be questioned. I hasten to add that in proposing to question the SP, I am in no way suggesting that we do away with symmetrization. Rather I will suggest that it can be understood as arising or falling from specific physical conditions that can indeed be elucidated. This is again a topic for future elaboration, but a starting point is the idea that a measurement interaction (suitably understood as involving a non-unitarity process) can break or initiate symmetrization, and that this corresponds to the advent or absence of distinguishability. This issue is discussed in Section 5. 

In any case, the Received View is exemplified by the following passage in Bigaj (2015) (emphasis added): \bigskip \

{\fontfamily{qcr}\selectfont

``The textbook way to introduce this postulate is through the concept of exchange degeneracy [Cohen-Tannoudji, C., Diu, B., $\&$ Laloe, F. (1977)].  Considering the joint state of two particles of the same type such that one of them occupies state $|u\rangle$ whereas the other one is in a different state $| v\rangle$, {\bf we should observe that the two permuted states $|u\rangle |v\rangle$ and $|v\rangle |u\rangle$ are empirically indistinguishable}. According to the essentialist approach this indistinguishability comes from the fact that both bi-partite states represent one and the same physical state of affairs. On the other hand, the haecceitist approach admits that there is a difference between the permuted and non-permuted states, but this difference cannot give rise to any observational effects, as haecceities are not empirically accessible. In order to avoid the degeneracy problem, we adopt the symmetrization postulate, which narrows down the admissible states to the symmetric (occupied by bosons) and antisymmetric ones (applicable to fermions).'' } \bigskip

The Received View (as exemplified in the above quote) can be questioned on (at least) two grounds: (1) it arguably errs in attributing ``empirical indistinguishability'' to two states that, according to the assertion of symmetrization as a postulate, {\it can never individually be empirically observed}; and (2)  it can be seen as neglecting some crucial physics. \smallskip

To elaborate regarding objection (1): This identifies in the received view a self-contradiction that (to my knowledge) has not been noticed. It pertains to the passage in (added) boldface in the above excerpt, which asserts (as a given) that ``the two permuted [direct product] states are empirically indistinguishable.''  However, the SP implies that the two permuted states $| u \rangle | v \rangle$ and  $|v\rangle |u\rangle$ are {\it never individually instantiated}, since it claims that identical quantum systems are always in symmetrized superpositions of direct product states. (In his more recent presentation of this argument, Bigaj (2022) uses the language ``by assumption'' to denote the assertion that the permuted states are empirically indistinguishable; but the problem remains that one cannot ``assume'' an empirical fact, especially about an entity claimed to be empirically inaccessible.) Thus the direct product states, according to the SP itself, are {\it never empirically observed}; and therefore they cannot be compared to see whether they are empirically indistinguishable.  (One might ask, in Humean fashion, ``From what experience or observation did this putative piece of empirical knowledge arise?'') Thus it cannot be claimed that the two states of affairs are empirically indistinguishable. Instead, the putative indistinguishability is an indirect metaphysical inference from phenomena that {\it are} empirically available, such as measurement statistics. 

To perhaps make this issue more clear, consider the usual situation of invariance under a transformation, such as rotation of a spherically symmetric body. We first empirically observe the body in some initial state $ | \Theta\rangle$. This means that the state has to be empirically accessible. Then we rotate it (subject it to the relevant transformation) and observe it in the transformed state
 $ | \Theta^{\prime}\rangle$, also empirically accessible. We observe that the body is unchanged (i.e., invariant) under this transformation connecting the initial and final states, since the two situations are empirically indistinguishable. We are justified in doing so because each of the states, individually, was empirically accessible. But the situation with the two permuted direct product states is crucially different, since the SP asserts that neither is ever empirically accessible. 
 \smallskip

Thus, I suggest that what has happened in the establishment of a consensus around the Received View---as exemplified by the Cohen-Tannoudji {\it et al} excerpt including the statement that two putatively non-instantiated states are ``empirically indistinguishable''---is that there has been a critical equivocation in the use of that term: in effect, it has been treated as equivalent to ``empirically inaccessible.'' But this is inappropriate. If a given set of states is empirically inaccessible, such states are not subject to the kind of inspection that would allow us to say that they are empirically indistinguishable under a given transformation, and thus involve a physical invariance, in contrast to the usual (legitimate) cases of invariance under transformations. And more importantly, in-principle observable quantities (such as probability densities) constructed from the product states are manifestly non-invariant under the permutation. 

A further concern is in order regarding the idea that there is a physical invariance (PI) attending the exchange permutation. There seems to be some equivocation in the literature in this regard, namely: in contrast to the claim that PI already attends the direct product states themselves (as in the Cohen-Tannoudji excerpt regarding ``exchange degeneracy''), it is sometimes presented as something that needs to be imposed on non-invariant mutually permuted product states. For example, Bigaj discusses this as a motivation for the SP (Bigaj 2022, 22-3), as do Eisberg and Resnick (1985, 305). (Thus Bigaj (2022) presents both redundancy and the need to impose invariance in support of the SP, although they would seem to be mutually exclusive--are direct product states invariant under the permutation, or are they not?) Specifically, product eigenfunctions are, in general, {\it non-invariant} under label permutation of identical quanta, and symmetrization is imposed to regain invariance of observable quantities (such as probability densities).\footnote{Bigaj doubles down on the idea that mathematically distinct states are nevertheless ``redundant'' in this passage: ``A well-known case of redundancy present in quantum mechanics is caused by the fact that two vectors that differ by a phase represent the same physical state.'' (Bigaj 2022, 54). However, this is falsified by such situations involving topological phase effects, as the Aharonov-Bohm effect. Cf. Robbins, (2003).} 

Thus, the term ``permutation invariance'' seems to be used in two disparate ways that often seem conflated in the literature. These are: (i) to characterize a putative invariance of the individual product states under a permutation of the indices; and (ii) to describe a requirement to {\it construct} states that are invariant (up to an overall phase) under permutation of the indices (which of course yields the symmetrized states). The EE interpretation, based on the idea that there is a redundancy of description involved in each of the direct product states under permutation, appears committed to (i), which we have already critiqued above. On the other hand, regarding (ii), if all one means by ``permutation invariance'' is that we don't already have it with the direct product states, but need to construct symmetrized states that {\it are} invariant under permutation, then of course such states are PI (but only up to a phase for the antisymmetric states) by construction. Then the issue is a different one: i.e.: ``{\it why} do we need to impose a form of permutation invariance on the physically relevant states?'', which is at least indirectly addressed in the current work. \smallskip

Now let us consider objection (2), that taking symmetrization as a postulate arising from a redundancy of description neglects specific physics:

In scattering situations involving identical quanta, there are distinct physical processes that must be equally represented in the theory and taken into account in a specific way. The key point is that these are distinct processes characterized by different topological structures, so that the ``redundancy of description'' traditionally invoked to support symmetrization (in the sense of (i) above) is not tenable here. An example of such an interaction is the electromagnetic repulsion between two electrons. This involves two kinds of processes or ``channels''. Each has a probability amplitude, and these amplitudes must be added (in a relativistic analog of symmetrization) in order to obtain the final amplitude for the overall interaction. These are depicted in Figure 1:

\vspace*{7pt}

\begin{figure}[!h]
\centering\includegraphics[width=1in]{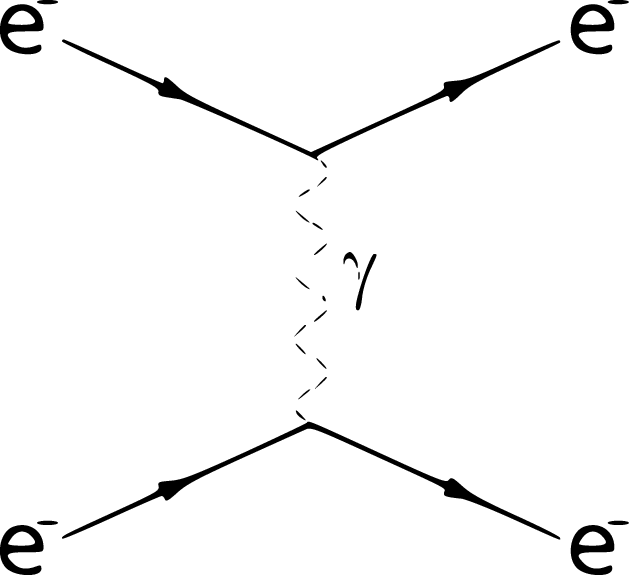}

\label{fig_sim}
\end{figure}
\begin{figure}[!h]
\centering\includegraphics[width=1in]{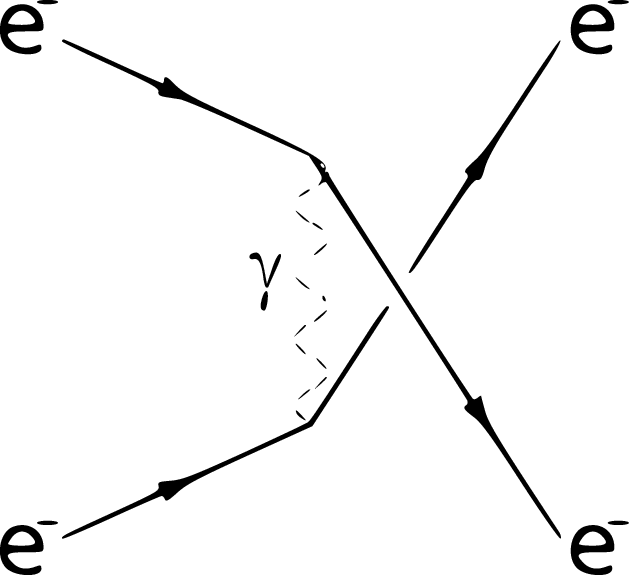}

\label{fig_sim}
\end{figure}

\vspace*{5pt}
\centerline{{\scriptsize Figure 1. The t-channel (top) and u-channel (bottom) for electron-electron (M\"{o}ller) scattering.}}\bigskip

Suppose we label the upper incoming electron as ``1'' and the lower incoming electron as ``2''.
  In words, the channels, as amplitudes, can be represented as follows:\bigskip

t-channel: ``amplitude for electron 1 to go up and electron 2 to go down''

u-channel: ``amplitude for electron 1 to go down and electron 2 to go up''\bigskip

(Here, the words ``up''  and ``down'' may be understood as just placeholders for outgoing momenta corresponding to specific detectors.) Importantly, there must be equal amplitudes of each process, since there is no physical reason for either to ``count'' more than the other. The channels are topologically distinct and therefore not just different names for the same situation. Therefore, symmetrization is not compelled by a redundancy of description, but by the requirement that both channels be equally present (with the phases demanded by the spin-character of the quanta).

 I hasten to add that these processes, as amplitudes, do {\it not} constitute ``facts of the matter'' about any trajectory; the electrons do not pursue a determinate trajectory under these conditions. This is a well-known feature of Feynman scattering amplitudes; i.e., that individual amplitude diagrams do not constitute ``facts of the matter'' about the careers of the described quanta. The channels may be understood as possibilities, not facts (as in measurement outcomes), and that is why they must be added before squaring to find the probability of the outgoing states, which are subject to measurement. 

Thus, in denying that the symmetrization of the scattering channels represents a redundancy of description, I am {\it not} claiming that the channels are {\it empirically distinguishable}; only that they have distinct physical referents. It should be noted that in order to evade this point, one must categorically deny that the expressions for the scattering channels have any physical referent; i.e., one must be committed to a strong form of antirealism/instrumentalism about quantum amplitudes.  For, if one even allows the possibility that the topologically distinct channels exist as physical processes (even if not empirically available), then clearly there is the possibility that ``redundancy of description'' does not obtain; and clearly one cannot invoke, as a reason for symmetrization, a premise that one allows may not obtain (although there might be other reasons). \newline

Interestingly, T. Bigaj (2022) considers this scattering situation, but he takes the different channels simply as formal depictions of the swapping of labels, rather than as real physical situations with distinct referents. Significantly, Bigaj's depiction of the two scattering channels obscures their differing topologies by showing neither of them as crossing. This is only possible by suppressing temporal evolution\footnote{So Bigaj's scattering illustration is not a Feynman-type diagram; even if those are not really spacetime diagrams, they show temporal order.} and disregarding the case when the four incoming and outgoing states are coplanar, so it is misleading in that respect. If one represents the processes in a Feynman diagram, the u-channel crosses while the t- channel does not.

 In any case, Bigaj uses the example to argue for what he terms a ``heterodox'' approach (Caulton's original term from his 2018), which involves dispensing with so-called ``factorist'' or direct-product states\footnote{Although both Caulton and Bigaj subscribe to the Received View as discussed above, which is apparently a much more firmly established form of orthodoxy than taking direct product states as physically meaningful.}). The present author, in questioning the SP, is certainly engaged in a heterodox approach \textemdash but clearly a very different one. So let us carefully distinguish these two ``heresies'' by terming Bigaj's approach {\it Heresy}$_{TB}$ and my own approach {\it Heresy}$_{RK}$. 

Nevertheless, we do have some common ground in that I certainly agree that for a case in which the spin states of the two fermions could serve as persistent labels for the spatial states, as in the state which I will call $|\Psi_D\rangle$ (Bigaj's eq (8.14):

 $$|\Psi_D\rangle = \frac{1}{\sqrt 2} [ | \uparrow_1 \rangle | L_1 \rangle  \otimes | \downarrow_2 \rangle | R_2   \rangle  -
 |\uparrow_2 \rangle | L_2 \rangle  \otimes | \downarrow_1 \rangle | R_1  \rangle ] \eqno(2) $$

\noindent  we have effective distinguishability of the two fermions, and that is why there is no interference in that case (provided the system is subject to a trivial Hamiltonian. However, real fermions would not behave this way; more on this below)).\footnote{Significantly, for the case with interference present, Bigaj looks only at the symmetrized spatial component of the state (his eqn. (8.10)), which implies that we are dealing with a case in which the collective spatial and spin degrees of freedom can be factorized. In this case, each such collective degree of freedom is subject to a symmetrized state. The latter form forbids the persistent ``marking'' of the individual quantum spins by their spatial states characteristic of
 $|\Psi_D\rangle$ (under a trivial Hamiltonian) and instead reflects spin-spin interaction, which is the basis of exchange forces. Bigaj's presentation obscures this subtle but crucial point.} \smallskip

Regarding (2), I would note that one would not need symmetrization in this case--it brings in superfluous content that drops out. Specifically, if (for the spin-aligned case with trivial Hamiltonian) we simply use a product state as in the case for the ``distinguishable'' particles (his equation 8.5), we obtain the same probabilities, as is already shown by Bigaj. But the matter can be put more strongly: under the given scenario of a trivial Hamiltonian, arguably symmetrization is unphysical. As alluded to above, imposing a trivial Hamiltonian on this example actually eliminates the exchange effects (sometimes called exchange ``forces'') corresponding to symmetrization. This crucial point has apparently not been noticed thus far, so it's worth elaborating. We turn to this issue in the next subsection.\smallskip
 
\subsection{Exchange effects require a perturbation} 

 The description of exchange effects through a specific Hamiltonian perturbation is well studied (e.g., Hutem and Boonchui, 2012; Stepanov, E. A. {\it et al}, 2018; Pavarini, E. {\it et al}, 2012), and is commonly termed ``Heisenberg exchange.''  Although it arises from the interchanging of identical quanta and not directly from a mediating force, the force-based (electromagnetic) interaction must still be present in order for the perturbation to be in effect.  The essential features of exchange forces can be understood as arising from a coupling of the spins of identical quanta, and its effect on the spatial part of the state. Specifically, the Heisenberg exchange Hamiltonian is: 
 
 $$ H_{Heis} = K_{ij} \langle \bf{s_i} \cdot \bf{s_j} \rangle    \eqno (3)  $$
 
where $K$ is the ``exchange integral''.

\smallskip For the case of electrons, $K$ arises from evaluating the expectation value, in the relevant symmetrized state, of the Coulomb interaction potential 

$$U=  \frac{e^2}{r_{12}} \eqno (4)$$ 

\noindent where $r_{12} = | \bf{r_2} - \bf{r_1}| $ is the separation of the quanta.
 
\smallskip Specifically, one finds two terms, one reflecting the Coulomb potential directly and the other, $K_{1,2}$, arising from exchange of the parameters associated with each quantum (cf. Hutem and Boonchui, 2012):
 
 $$K_{1,2} = \int_{\bf{r_1}} \int_{\bf{r_2}}  \psi^*_a(\bf{r_1})  \psi^*_b(\bf{r_2})  \frac{e^2}{| \bf{r_2} - \bf{r_1}|} 
    \psi_a (\bf{r_2})  \psi_b (\bf{r_1})  d\bf{r_1} d\bf{r_2} \eqno(5)  $$
 
 Although this example makes use of the Coulomb potential, the electromagnetic basis of the exchange interaction is still applicable for the case of neutral fermions, since they have a magnetic moment.\footnote{This may be confirmed by noting that the Coulomb potential may be replaced by the magnetic dipole potential in (4).} Thus, if no field is present, then $K$ is zero and there can be no exchange interaction. Equivalently, if the spins are stipulated to be fixed in the individual state spaces throughout a process, which denies the coupling in (3), then the exchange interaction is suppressed. \smallskip
 
 We therefore see that the interaction corresponding to the exchange ``force'' is not permitted under the trivial Hamiltonian used in Bigaj's treatment (or in any treatment that simply looks at a prepared symmetrized state without taking into account the relevant exchange perturbation).\footnote{Indeed, such states are eigenstates of the unperturbed Hamiltonian with energy degeneracy (simply the sum of the individual electron energy eigenvalues, independent of spin orientation; cf. Koch (2012) in Pavarini et al (2012), eqn. (7). In order to gain lifting of the degeneracy and dependence on spin orientation characteristic of exchange interaction, the perturbing electromagnetic Hamiltonian is required. Expecting to have exchange effects without this perturbation (by writing down a Slater determinant) is like expecting excited atoms to decay in stationary bound states of the unperturbed Hamiltonian; it won't happen.}  If the field-based perturbation is omitted, then symmetrization is superfluous, since the exchange-interaction part of the Hamiltonian will be zero without it, regardless of indistinguishability of the quanta.  \smallskip
 
Elementary textbook descriptions of exchange forces often use factorized space and spin states, and in this idealized form one can look only at the spatial state to help visualize exchange effects. This makes it appear as though such effects result only from the {\it form} of the state itself and not any perturbative Hamiltonian. But in this case, it is still the spin-based nature of the quanta that enforces the symmetrization, and spin has a magnetic moment; thus for real quanta, the perturbation is always present. Eliminating it eliminates the exchange effects, and thus arguably makes the symmetrization inapplicable, at least insofar as symmetrization is expected to correspond to exchange effects (which are forms of entanglement correlations).\footnote{Treating all identical quanta as always in symmetrized states without taking into account the exchange perturbation arguably leads to problematic issues such as lack of entanglement in what looks like entangled states, so that research on ``what counts as entanglement?'' then seems to be in order. An example of (in my view, needless) complications arising from imposing symmetrization universally, without the exchange perturbation in effect, and observing that this doesn't correspond to `real' entanglement may be seen in Plastino, Mansano and Dehesa (2010). This is perhaps a cautionary reminder regarding mathematical overgeneralization of theoretical quantities, at the expense of vital physics.} 

\smallskip

This point is further reinforced by an interesting result by Jabs (2010). He provides a simple model connecting spin to statistics by imposing a consistency condition on the manner in which spin transforms under rotation by an azimuthal angle $\chi$ around a common spin quantization axis. It should be empasized that the latter is a collective feature of the multiquantum system, which Jabs terms a ``spin frame''.\footnote{The spin quantization axis need not coincide with the z-axis of the spatial wave function, so he gives it a distinct parameter $\chi$.} Specifically, he demands that both (all) spins transform in the same way, where the transformation is implemented through phase factors in a product of two (or more) states. Imposing this consistency condition yields the specific symmetrized forms in a simple manner (which, at least to this author, suggests that Jabs' model reveals a deep physical basis for the symmetrized forms). The definition of a common spin quantization axis, where such an axis only becomes relevant in the presence of a field, reinforces the role of the field associated with spin (even in the case of neutral fermions). It also implies that if there is no possibility of the spins interacting, then there is no need to take into account a common spin frame, and thus no need for the symmetrization (which, as Jabs notes, arises from the capability of interference between the amplitudes for the mutually permuted states).

Jabs' model also suggests that it is the rotation group, not the permutation group, that underlies the relevant symmetry involving exchange of indices. This would also explain why we never see ``parastatistics'' in nature, and also reinforces the concern expressed above that there is not really an applicable permutation symmetry involved (since there is no {\it empirically observable} permutation invariance at the level of the individual states, and there is at the very least mathematical non-invariance of product state eigenfunctions under permutation). Nevertheless, it should be noted that the Jabs model is not required for the points raised in the current work. In particular, it may be noted that $H_{Heis}$ (3) has rotational invariance. This, along with the non-invariance of product wavefunctions under permutation, also suggests that it is rotational invariance, rather than permutation invariance, that is physically applicable to the exchange.

\smallskip
\section{Exchange of essences inadequate to support the physics} 

In section 1d, we discussed two distinct ways of interpreting the transposition of labels or indices appearing in the symmetrized states: either (1) Exchange of Essences (EE) or (2) Exchange of Haecceities (EH). In this section, we will see why EE falls short of accounting for the form of the symmetrized states. 

	Exchanging essences amounts to the idea that when we exchange the labels attributed to the quanta, we are exchanging their essences (i.e., essential properties). This would change nothing about the physical situation, since the essences are identical. As noted above, this approach attributes a property to the collective system called ``permutation symmetry'' or ``permutation invariance,'' since EE implies that the physical situation is completely unchanged by the permutation of the labels. This means, essentially, that the two states in the superposition are simply two names for the same thing (a form of representational redundancy). And this is where the EE approach gets into trouble. Of course, we have already questioned the idea that there is truly an invariance with respect to permutation, but there is a further problem with assuming this sort of redundancy of description, as follows: 

Consider the planet Venus. It has two names that refer to it equally well: ``Morning Star'' (MS) and ``Evening Star'' (ES). While the names look different, they refer to the same physical object. But this means that it does not matter {\it how much} of each name we use. That is, any state of the form

$$ a |MS\rangle  + b |ES\rangle \eqno(6) $$

\noindent refers to Venus, as long as a and b satisfy the normalization condition (which, for quantum theory, requires that their squares add up to unity). Invoking just ``Morning Star'' gives us Venus; invoking just ``Evening Star'' gives us the same Venus; invoking any other combination (suitably normalized) gives us Venus. Thus, treating permuted states like $A(x_1)B(x_2)$ and $B(x_1)A(x_2)$ as simply two different names for the same thing---a consequence of EE---fails to provide us with the necessary structure of the symmetrized states that are clearly required for empirical correspondence. Specifically, we must have both permuted states in equal amounts. Helping ourselves to the empirically required superposition is then not so much a ``postulate'' as a severely {\it ad hoc}  move, since clearly, according to EE, one or the other product state should suffice. 

The vital physical role of each of the permuted states becomes even more explicit when we consider scattering process such as that discussed in Section 2a, which cannot physically take place without both channels. Each of the states in the superposition corresponds to a different channel, and they must both be taken into account in equal magnitudes. Thus, the scattering case serves as a counterexample to the idea that permutation represents a redundancy reflecting an invariance. The situation is clearly not invariant under the permutation, since it corresponds to distinct scattering processes with differing topological structures. As we noted in the previous section, one would need to categorically deny that the scattering channel amplitudes refer to anything real in order to evade this point.

Concerning the non-empirical nature of the product states representing, in this case, scattering channels, Bigaj says: ``Operators which represent properties of individual particles are meaningful but, strangely enough, they are not literally observables.'' (Bigaj 2015, 60) 
 The only reason this seems ``strange'' is because, contrary to our usual classical expectations, there is no fact of the matter about which channel is in play---they are both in play. Thus, there is no fact of the matter about the contingent ``properties of individual particles'' in this context, and that is why such operators do not correspond to empirically observable properties. But again, the fact that channels are not {\it observationally distinct} does not logically imply that they are not real, distinct physical processes. And if they are real, distinct physical processes, then it cannot be asserted that the scattering amplitudes are mutually redundant expressions. Other examples of physically meaningful operators that do not correspond to empirically observable quantities are field creation and annihilation operators. Clearly, these are efficacious and consequential; fields are being created and annihilated by way of them, and fields are detectable. So again, one would have to embrace explicit antirealism about physical theory (and for consistency, deny that field creation and annihilation operators represent anything in the world) to maintain that non-empirically available theoretical entities (such as the scattering channel amplitudes) have no physical referent and therefore can be considered redundant and interchangeable (as required by EE).
Thus, the conclusion is that EE does not provide sufficient structure to support the symmetrized states. In stronger terms, it is arguably falsified by the relevant physics, which involves distinct processes (provided one allows for the possibility that the theoretical quantities refer to entities and processes in the world) and demands that both direct product states be represented in equal amounts.

\section{Quantum Haecceity} 

	The above considerations remind us that the labels attributed to the quanta in product states such as  $|A\rangle_1 |B\rangle_2$  can be understood as having distinct referents, since exchanging the labels can be seen as changing the physical situation. This leads us to the conclusion that it is their haecceities, not their identical essences, that are being exchanged when the labels are permuted. However, clearly the quanta do not qualify as independent individuals in the usual classical sense, since we have exchange effects that result from a ``blending'' of these putative haecceities (quantitatively described by $H_{Heis}$, which demands symmetrization). Thus, we cannot be dealing with a classical sort of haecceity implying full-fledged individuality. Instead, we have what I would like to call {\it quantum haecceity} (QH), which involves a form of potentiality. While the latter notion is obviously a matter for further study, as a starting point we could note that a composite state in the form of a wavefunction $\Psi(x_1, x_2)$ such as
	
	$$  \Psi(x_1, x_2) = \frac{1}{\sqrt 2} [  \psi(x_1)\phi(x_2) + \psi(x_2) \phi(x_1)  ] \eqno (7)$$
	
\noindent suggests that the indices in parameters $x_1$ and $x_2$ represent the potential for two distinct outcomes upon performing a measurement of position; whereas if there were only one quantum, we would have only one outcome.  The labels or indices in (7) signify potentialities rather than actualities because there is no fact of the matter about any actual possession of a position-property by either of the quanta described by this state. Thus, in more quantitative terms, if the cardinality of the entanglement is N  (i.e., we have N entangled systems), then we must have indices $i \in \{1,N\}$ to represent the set of possible outcomes ${o^j}_i,  i \in \{1,N\}, j \le N $   corresponding to observable(s) $O^j$ measured locally on each of the quanta.\footnote{Krause and French (2007) discuss cardinality in connection with quantum systems and conclude that cardinality alone does not confer full individuality. The present study does not contradict their finding, since quantum systems under conditions involving symmetrization have only a quasi-individuality, in keeping with the weakening of haecceity accompanying QH. It is important to notice, however, that cardinality alone is not the only feature that enters into QH; also relevant is the possibility of distinguishing measurement, as discussed in relation to (7).} 

	Another, more qualitative way to get a sense of this notion of QH is in terms of organic systems. It is well known that trees of the same type, originating from distinct seeds, can merge into effectively a single entity (see Figure 2). 
	
\begin{figure}[!h]
\centering\includegraphics[width=3in]{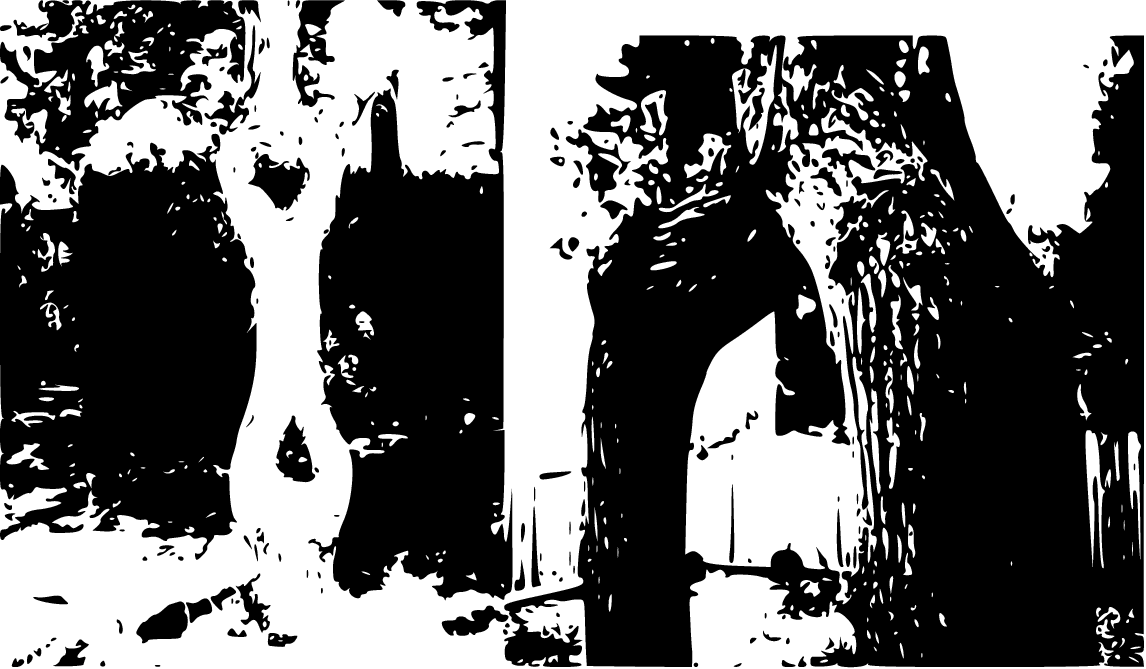}

\label{fig_sim}
\end{figure}
\newpage
\centerline{{\scriptsize Figure 2. The organic quality of quantum haecceity.}}
\bigskip

 The trees shown in Figure 2 have been deliberately coaxed into the forms shown, but trees in nature also naturally merge and separate given the appropriate conditions. The analogy with our current topic is that the saplings share identical essences but differing haecceities corresponding to their cardinality: specifically, there are N saplings. On the left, two saplings have been planted and coaxed into joining to form a collective entity of the same essence, with the potential to diverge and reunite again, which has occurred. Without the haecceities corresponding to their number, they would have no potential to converge and to diverge as shown, since there would be only a single entity. 

 	The above situation mirrors the behavior of correlated quantum systems. A symmetrized state with a form like (7) represents a collective whole with the potential for separation into distinct property-states, with no fact of the matter about ``which system would go where'' if such separation were to occur. The indices represent only the potential for two differing outcomes, given an appropriate measurement interaction.  Thus, quantum haecceity is a form of haecceity that corresponds to the cardinality of the systems available for measurement. It does not confer upon those systems a full-blown individuality, but merely represents the potential for differing outcomes upon measurement. 
	
	 Interestingly, Jabs' model leads to a distinction between exchange of quantum label indices and exchange of function (e.g., wavefunction) parameters. It indicates that it is not the quantum labels themselves that are exchanged, but rather the function parameters. In effect, quantum-labels become redundant. Since parameters such as $x_1$ and $x_2$ represent possible outcomes for two measurements of position, this reinforces the idea that the indices reflect the potential for various outcomes, given by the cardinality of the correlated system. We consider this concept in somewhat more detail in the next section.

\section{Symmetrization and non-symmetrization}  

\subsection{Measurement and distinguishability}  

A further implication of the foregoing is that the measurement process results in distinguishability of the measured systems, since upon the occurrence of an outcome, they possess distinct properties corresponding to the eigenvalues of the measured observables. One can say that specific potentialities have been elevated to actualities, which now serve to distinguish the quanta (corresponding to the emergence of ``classicality''). But the labels themselves play no part in this distinguishability, since there is still no matter of fact about which system has which property. Rather, their correlation has been broken through actualization of measurement results, so at this point (as mentioned in the previous section) the labels are revealed as surrogates for the distinct measurement outcomes represented by function parameters. 

For example, a photon absorbed by the detector on the right is no longer correlated with the photon absorbed by the detector on the left. Or, two initially correlated electrons may be individually detected by becoming bound to two different atoms. In either case, the exchange correlations are broken, and the systems are no longer (even formally) entangled. They can now be distinguished by their detected properties, even if not by their indices. The indices, which index quantum haecceties (QH), reflected the cardinality of the entanglement, which in turn represents the potential for the number of outcomes upon measurement.\footnote{This approach addresses Bigaj's question: ``How do we know that a particular system of electrons consists of five, ten or $10^{12}$ electrons? How can we count them, if we cannot distinguish them in any way?'' Of course, ontology is not epistemology, and a system's cardinality does not depend on our knowledge. But still, the cardinality of the system is reflected in its potential for a given number of measurement outcomes.} 
Upon measurement, the QH are transformed into specific distinguishable, contingent properties. This again reflects their status as indexing potentialities, not actualities. 

Thus, in order to allow for the concept of QH, one must allow that quantum theory describes potentialities and their elevation into actualities through the process of measurement. The present author recognizes that this is perhaps an unfamiliar and even radical notion, although it certainly has longstanding precedent. Examples in the literature are: Heisenberg (1955, 12); Shimony (1997); Toraldo di Francia (1976); Kastner, Kauffman, and Epperson (2018); Jaeger (2017). Howard (1989) also notes that providing a measurement context allows for this sort of actualization, although he doesn't explore the matter in detail. However, his suggestion may be considered a precursor, at least in spirit, to the idea presented here. 

	The above considerations bring us back to the measurement problem of standard quantum mechanics, which is solved in the transactional approach (most recently updated in its fully relativistic form as RTI; cf. Kastner 2018, 2020, 2022). What I mean by ``standard quantum mechanics'' is the idea that there is no real, physical non-unitarity in quantum theory (and at most a ``projection postulate'' with no corresponding physical account).\footnote{This situation leads to well-known problems such as Schr\"{o}dinger's Cat and the Wigner's Friend paradox.}  While we do not have the space to go into detail here regarding the transactional formulation, the interested reader may consult the references provided herein. For present purposes, we may note that RTI provides a rigorous account of the advent of the non-unitary measurement interaction, which projects an entangled or symmetrized system into a physically relevant factorized state. For example, measurement of spin along $\bf{z}$ of electrons prepared in a spin singlet state results in the transition: \bigskip
	
	$$ \frac{1}{\sqrt 2} (| \uparrow\rangle_1 | \downarrow\rangle_2  -  | \downarrow\rangle_1 | \uparrow\rangle_2) \rightarrow 
	| \uparrow_{\bf{z}} \rangle_1|  \downarrow_{\bf{z}} \rangle_2    \eqno (8) $$

	\bigskip
	
\noindent where the right-hand side denotes two electrons that are now distinguished by their respective measurement results. Again, at this point the indices `1', `2'  are merely surrogates for the spatial component of the state corresponding to leftward or rightward momentum. The measured electrons can (and should be) described by a direct product state in this picture, because such detected electrons are prevented from participating in a mutual exchange interaction and are truly (not just approximately) distinguishable by virtue of the non-unitary measurement transition. This is because in order to be detected, a fermion must become part of a bound state. We address this issue more fully in Section 5.b. 

	While some authors exploring the EE approach have suggested that the measurement problem is not at issue in discussions of quantum metaphysics (addressing such topics as individuality in quantum systems), in fact it does appear to be a motivation for work by A. Caulton, who rejects what he calls the ``factorist'' approach.\footnote{T. Bigaj subscribes to Caulton's program, noting in his (2022) that Caulton (2018) "became a bible for me'' (Bigaj 2022, vii).}  He says: ``One consequence of the non-individuality of factorist systems is that they cannot become classical particles in the appropriate limit, since [they] must remain in statistically mixed states.'' (Caulton, 2018). Dieks and Lubberdink (2022) also note that ``it is a consequence of the standard view that identical quantum particles remain indistinguishable even in the classical limit, which makes a smooth transition to the classical particle concept impossible'' (Abstract).

	However, as allowed by Dieks and Lubberdink, that concern applies only to the standard theory lacking a specific measurement interaction. Under the transactional formulation, quantum systems can indeed be distinguished by their outcomes and can therefore approach classicality upon instantiating a measurement outcome. The applicable mixed states are then proper mixtures (rather than irreducibly improper) and are simply epistemic tools reflecting our ignorance of the actual outcomes (cf. Kastner, 2020), which do indeed serve to confer at least pseudo-classical distinguishability.\footnote{We say ``pseudo'' here since such systems are still subject to quantum processes as spreading of wave packets and interactions that can re-entangle them.} (This brings us back to Howard's proposal, albeit with a physically grounded reason for taking seriously the mixed states.)
	
	Thus, Caulton's program appears ultimately motivated by the standard theory's inability to define the conditions for measurement and thus its inability to obtain classical-level distinguishability from specific physics. This problem is remedied in the transactional formulation, under which the labels can be understood as representing the potentialities for specific numbers of measurement outcomes corresponding to the same number of measured observables. It should be emphasized that TI is empirically equivalent at the level of the Born probabilities to the standard theory (since TI in fact derives the Born rule) and does not change the theory in an {\it ad hoc} manner. Specifically, TI does not change the Schr\"{o}dinger equation (as in other ``spontaneous collapse'' approaches). It simply takes into account that fields operate according to the direct-action theory, which involves absorber response (and thus non-unitarity) under well-defined conditions (see Kastner 2022, Chapter 5, for quantitative specifics). Physical non-unitarity is missing in the standard approach, which presupposes (as a basic metaphysical commitment) a theory of fields in which propagation is unilateral and always unitary. (Questioning that particular metaphysical commitment is another aspect of $\it Heresy_{RK}$.) 
	
\subsection{Physical conditions for symmetrization} 

 I have argued in the foregoing that symmetrization is required, or not required, under specific physical conditions, and thus need not (and should not) be regarded as a universal postulate. This of course raises the question: precisely what are the conditions for requiring symmetrization and for dropping it? In the preceding section, I noted that measurement outcomes obtained in a non-unitary physical transformation warrant dropping symmetrization, and that in that case unsymmetrized product states (involving eigenstates of the measured observables) correctly reflect the physical situation.  \smallskip

 More generally, it can be argued that symmetrization of identical quanta applies only to (i) free quanta and (ii) quanta subject to a common bound state.\footnote{The present treatment addresses fermions in particular. Physical symmetrization requirements for bosons, such as photons, are a somewhat different matter. But they still involve a nontrivial interaction Hamiltonian (cf.  Franson 2004). The question of whether photons have a magnetic moment is a fascinating one; Saglam and Sahin (2015) argue that they do.}  An example of (i) is a pair of correlated electrons prepared in a spin-singlet state allowed to diverge, as in a typical Bell-type experiment; an example of (ii) is a pair of electrons in an ortho- or parahelium state.    \smallskip

Let us consider in more detail the conditions for dropping symmetrization. This also involves ``getting our hands dirty'' by looking at the details of detection, which are traditionally glossed (arguably because of the measurement problem attending the standard approach).  In particular, detection of charged fermions involves such processes as detecting a current. This means that new charges (e.g., electrons) have entered the system generating the current, which in turn means they have entered a bound state (such as a conduction band).  Another example of detection of an electron would be radiative recombination, in which an atomic ion captures an electron, which would render the atom neutral and could be detected as a drop in current. In both cases, the electron must radiate (a non-unitary process in the transactional formulation) in order to enter a bound state.\smallskip 

On the other hand, electrons could become subject to symmetrization by leaving a bound state, for example through the photoelectric effect (the inverse of radiative recombination), in which the electron absorbs a photon. Such an electron would then become subject to symmetrization, since it is no longer shielded from the exchange interaction.  Thus, the fact that fermions always must radiate or absorb to enter or leave a bound state (a non-unitary process in the transactional formulation) is why this kind of transition heralds the advent or departure of symmetrization.

 In this regard, it should be recalled that bound states are poles in the scattering matrix, which already suggests a breakdown of unitarity (see Kastner, 2017).  Extant discussions that omit $H_{Heis}$ and treat symmetrization as universal  suppress the bound state nature of many orthogonal spatial states (at least those applying to controlled localization), and thus implicitly misrepresent controllably localized quanta as free quanta. Figure 3(a) (below) shows the physically accurate situation in which detected quanta are separated by potential barriers and thus distinguished; in this situation, the exchange interaction is prohibited, and arguably symmetrization is inappropriate. On the other hand, 3(b) shows the picture implied by accounts that disregard the physically bound status of detected or controllably localized quanta. In 3(b), the spatial states are nominally orthogonal, but since these apply to free quanta, they are not shielded from $H_{Heis}$. The latter, not a postulate, explains why a symmetrized state applies in such a situation; yet it is routinely neglected. Many extant accounts attribute exchange effects only to the form of the assumed-universal symmetrized state, and ignore $H_{Heis}$, which makes the separability of states like (2) mysterious. And as discussed in Section 2.b, in fact they {\it are} separable without $H_{Heis}$, but are also unphysical in that case.
 \bigskip	
\begin{figure}[h!]
  \renewcommand{\thefigure}{Z}
  \begin{center}
    \begin{tikzpicture}
      \draw [thick] (0,0) 
      to [out=-10, in=170] (.5,-1)
       to [out=10, in=190] (1,0)
      to [out=0, in=180] (3,0)
      to [out=-10, in=170] (3.5, -1)
      to  [out=10, in=190] (4,0);    
      \draw (0.5,-.5) node [black,above] {X};
      \draw (3.5, -.5) node [black,above] {X};
      \draw (2,-1) node [black,below] {3(a)};
      \draw [thick] (0,-4)
      to [out=0, in=180](4,-4);
       \draw (0.5,-4) node [black,above] {X};
      \draw (3.5, -4) node [black,above] {X};
      \draw (2,-5) node [black,below] {3(b)};
    \end{tikzpicture}
  \end{center}
\end{figure}

\centerline{\scriptsize{Figure 3. Fermions: bound (a) vs. non-bound (b).}}
\bigskip
\section{Conclusion} 

	I have argued that the proper way to understand the permutation of labels or indices in multi-quantum correlated states for indistinguishable quanta is in terms of the exchange of haecceities (EH) rather than the exchange of essences (EE). EE asserts that the permutation does not change the physical situation, while arguably the physical situation does change under the permutation, as discussed herein.  A case in point is the existence of two topologically distinct scattering channels for the electron-electron repulsion interaction, which correspond to the two different states of the permutation--both being required in equal amounts. Since EE considers the mutually permuted product states as simply different names for the same physical situation, it provides no physical reason for equal amplitudes of both components, as required for symmetrization consistent with empirical results (as illustrated by the ``Morning Star'' and ``Evening Star'' example).
	
	However, the usual notion of haecceity does not apply to quantum systems, since it corresponds to full individuality. The blending of identities of the quantum systems in symmetrized states indicates that they are not full-fledged individuals. Thus, I propose that we need a new concept of haecceity applying to quantum systems: quantum haecceity (QH), which reflects the cardinality $N$ of entanglement and does not correspond to full, classical individuation. A natural way to interpret this is in terms of the potential for a measurement outcome: i.e., $QH_i$ represents the potential for an outcome ${o^j}_i,  i \in \{1,N\}, j \le N $ upon measurement of an observable $O^j$ (where $j$ can be freely chosen; the set of observables need not commute). This is reminiscent of proposals by Howard (1989) and Toraldo di Francia (1976).
	
	 I have pointed out that ``measurement'' becomes physically well-defined under the Transactional Interpretation (particularly in its relativistic form, RTI, as elaborated in Kastner, 2018, 2020, 2022). In the transactional formulation, quantum systems can indeed become distinguished by their measurement outcomes, thus instantiating an approach to classicality under which a simple product state correctly represents the physics, even if only temporarily (until, for example a bound system is liberated and thus able to interact with other identical systems again and participate in exchange effects). Thus, one need not deny the physical relevance of the direct product states. On the contrary, they may be seen as meaningful because they not only represent crucial physics that supports symmetrization where it is appropriate, but they also reflect the existence of distinguishable measurement outcomes where these are empirically present.
		
	 I have also pointed to a problem in assuming that there is a true physical invariance under permutation, since neither of the mutually permuted direct product states appearing in a symmetrized state is empirically accessible. Moreover, individual product state wavefunctions (amplitudes) are manifestly non-invariant under the permutation. Thus, the usual (valid) situation warranting the identification of an invariance or symmetry transformation, in which a system remains unchanged under a transformation from one state to another transformed state, is not instantiated for the direct product states. Instead, an invariance is traditionally assumed {\it post hoc} based on the empirically required symmetrized states. These states can be seen as possessing symmetry, but it is more accurately seen as rotational symmetry, based on the form of the exchange interaction (Heisenberg Hamiltonian), than permutation symmetry.
	 
	  In addition, I have noted that exchange forces require that a perturbative field be present and that spins be allowed to interact, as described by the Heisenberg Hamiltonian. In the absence of such conditions (for example in fermions bound in separate potential wells, or in hypothetical cases of a trivial Hamiltonian), the physics underlying quantum state symmetrization is absent and thus under these conditions symmetrization is, at the very least, superfluous.

\ack{I am grateful for valuable correspondence with Federico Holik, useful comments from two anonymous referees, and a helpful critical reading of an earlier version of the manuscript by Edouard Machery.}

\section{References}

Bigaj, T. (2015). ``Exchanging Quantum Particles,'' {\it Philosophia Scientiae 2015}:(19-1),185-198).

Bigaj, T. (2022). {\it Identify and Indiscernibility in Quantum Mechanics}. Cham: Palgrave-MacMillan. 

Bigaj, T. and Ladyman, J. (2010). ``The Principle of the Identity of Indiscernibles and Quantum
Mechanics,'' {\it Philosophy of Science 77}, pp. 117 - 136.
DOI: https://doi.org/10.1086/650211

Caulton, A. (2018): ``Qualitative Individuation in Permutation Invariant Quantum Mechanics,''
arXiv:1409.0247v1.

Chakravartty, Anjan, "Scientific Realism", The Stanford Encyclopedia of Philosophy (Summer 2017 Edition), Edward N. Zalta (ed.), URL = <https://plato.stanford.edu/archives/sum2017/entries/scientific-realism/>.

Cohen-Tannoudji, C., Diu, B., \& Laloe, F. (1977). Quantum Mechanics. London, Paris: Wiley,
Hermann.

Dieks, D. and Lubberdink, A. (2022). ``Identical Quantum Particles as Distinguishable Objects,'' {\it Journal for General Philosophy of Science / Zeitschrift f\:{u}r Allgemeine Wissenschaftstheorie 53}, (3):259-274.

Eisberg, R. and R. Resnick (1985). {\it Quantum Physics of Atoms, Molecules, Solids, Nuclei, and Particles}. New York: John Wiley \& Sons. 

Franson, J.D. (2004). ``Comment on Photon Exchange Interactions.'' https://arxiv.org/pdf/quant-ph/0405141.pdf.

French, S. and Redhead, M. 1988, ``Quantum Physics and the Identity of
Indiscernibles,'' {\it British Journal of the Philosophy of Science, 39}: 233?46.

French, S., 2019, ``Identity and Individuality in Quantum Theory,'' The Stanford
Encyclopedia of Philosophy (Winter 2019 Edition), Edward N. Zalta.

Heisenberg, W.	(1955). ``The	Development	of	the	Interpretation	of	Quantum	Theory,''	
in {\it Niels Bohr and the	Development of	 Physics}. ed. Wolfgang Pauli (NewYork: McGraw-Hill).

Howard, D. (1989). ``Holism, Separability, and the Metaphysical Implications of the Bell Experiments,'' in Cushing, J. and E. McMullin, Eds., {\it Philosophical Consequences of Quantum Theory: Reflections on Bell's Theorem}. Notre Dame: University of Notre Dame Press (1989).

Hutem, A. and S. Boonchui (2012). ``Evaluation of Coulomb and exchange integrals
for higher excited states of helium atom by using
spherical harmonics series,'' {J Math Chem 50}:2086-2102.
DOI 10.1007/s10910-012-9997-6

Jabs, A. (2010). Connecting Spin and Statistics in Quantum Mechanics. {\it Found Phys 40}: 776-792. https://doi.org/10.1007/s10701-009-9351-4

Jaeger, G. (2017). ``Quantum Potentiality Revisited,'' {\it Phil. Trans. R. Soc. A 375}: 20160390.

Kastner, R. E. (2017). ``Bound states as fundamental quantum structures.'' In: {\it Quantum Structural Studies}, ed. Jekni\'k-Dugi\'c, J., Jaroszkiewicz, G. and Kastner, R. E. , Singapore: World Scientific (2017), pp. 427-432. 
doi: 10.1142/9781786341419\_0013. 

Kastner, R. E. (2018). ``On the Status of the Measurement Problem: Recalling the
Relativistic Transactional Interpretation,'' {\it Int'l Jour. Quan. Foundations 4}, 1: 128-141.

Kastner, R. E. (2020). ``Decoherence and the Transactional Interpretation,'' {International
Journal of Quantum Foundations 6}, 2:24-39.

Kastner, R. E. (2022). {\it The Transactional Interpretation of Quantum Mechanics: A
Relativistic Treatment}. (2nd Ed.) Cambridge: Cambridge University Press.

Kastner, R.E., S. Kauffman, M. Epperson (2018). ``Taking Heisenberg's Potentia Seriously,''
{\it International Journal of Quantum Foundations 4}:2, 158-172.

Koch, E. (2012). ``Exchange mechanisms," in Pavarini et al (2012), Chapter 7.

Krause, D. and French, S. (2007). ``Quantum sortal predicates.'' Synthese 154 (3):417 - 430.

Loemker, L., 1969, (ed. and trans.), {\it G. W. Leibniz: Philosophical Papers and Letters}. 2nd
ed., Dordrecht: D. Reidel.

McMullin, E. (1984). ``A Case For Scientific Realism,'' In J. Leplin (ed.) {\it Scientific Realism}. Berkeley, CA: University of California Press.

Motl, L. (2012) https://physics.stackexchange.com/questions/39686/entangled-electron-positron-pair. Accessed 2.1.2023.
 
Pavarini, E., E. Koch, F. Anders, and M. Jarrell (2012). {\it Correlated Electrons: From Models to Materials.} Modeling and Simulation, Vol. 2. Forschungszentrum Julich, ISBN 978-3-89336-796-2. http://www.cond-mat.de/events/correl12

Plastino, A.R., D. Manzano, and J. S. Dehesa (2010).  ``Separability Criteria and Entanglement Measures for Pure States of N Identical Fermions,'' https://arxiv.org/pdf/1002.0465.pdf

Robbins, Jonathan M.. ``Topological Phase Effects'' (2003). {\it Digital Encyclopedia of Applied Physics.}  Wiley. https://doi.org/10.1002/3527600434.eap525

Saglam, Z. and Sahin, G. (2015). ``Magnetic Moment of Photon,''
{\it Journal of Modern Physics 06(07)}:937-947. DOI: 10.4236/jmp.2015.67098.

 Shimony,	A.	(1997)	"On	Mentality,	Quantum	Mechanics, and	the	Actualization	of	
Potentialities,"	in	R.	Penrose,	{\it The	Large,	the	Small	and	the	Human
Mind}. Cambridge	UK:	Cambridge	University	Press,	1997),	pp.	144-160.

Stepanov, E. A. {\it et al} (2018). ``Effective Heisenberg model and exchange interaction for strongly correlated systems,''
 {\it Phys. Rev. Lett. 121}, 037204. (https://arxiv.org/pdf/1802.10068.pdf)
 
 Toraldo di Francia, G. (1976). {\it The Investigation of the Physical World,} Cambridge: Cambridge University Press.





\end{document}